\documentclass[%
aps,
 amsmath,
 amssymb, 
reprint,%
numerical,
]{revtex4-1}
\usepackage{epstopdf} 
\usepackage{color}
\usepackage[english]{babel}
\usepackage{amsmath}
\usepackage{amsfonts}
\usepackage{graphicx}
\usepackage{array}

 \makeatletter
  \providecommand\color[2][]{%
    \GenericError{(gnuplot) \space\space\space\@spaces}{%
      Package color not loaded in conjunction with
      terminal option `colourtext'%
    }{See the gnuplot documentation for explanation.%
    }{Either use 'blacktext' in gnuplot or load the package
      color.sty in LaTeX.}%
    \renewcommand\color[2][]{}%
  }%
  \providecommand\includegraphics[2][]{%
    \GenericError{(gnuplot) \space\space\space\@spaces}{%
      Package graphicx or graphics not loaded%
    }{See the gnuplot documentation for explanation.%
    }{The gnuplot epslatex terminal needs graphicx.sty or graphics.sty.}%
    \renewcommand\includegraphics[2][]{}%
  }%
  \providecommand\rotatebox[2]{#2}%
  \@ifundefined{ifGPcolor}{%
    \newif\ifGPcolor
    \GPcolorfalse
  }{}%
  \@ifundefined{ifGPblacktext}{%
    \newif\ifGPblacktext
    \GPblacktexttrue
  }{}%
  \let\gplgaddtomacro\g@addto@macro

  \makeatother
 
\begin{document}

\title{Appearance of effective surface conductivity -- an experimental and analytic study}
\author{Jakub Lis}
\email{j.lis@uj.edu.pl}
\affiliation{Center for Nanometer-Scale Science and Advanced Materials (NANOSAM), Faculty of Physics, Astronomy and Applied Computer Science, Jagiellonian University, ul. St.  Lojasiewicza 11, 30-348 Krakow, Poland}
\author{Mateusz Wojtaszek}
\affiliation{Center for Nanometer-Scale Science and Advanced Materials (NANOSAM), Faculty of Physics, Astronomy and Applied Computer Science, Jagiellonian University, ul. St.  Lojasiewicza 11, 30-348 Krakow, Poland}
\author{Rafal Zuzak}
\affiliation{Center for Nanometer-Scale Science and Advanced Materials (NANOSAM), Faculty of Physics, Astronomy and Applied Computer Science, Jagiellonian University, ul. St.  Lojasiewicza 11, 30-348 Krakow, Poland}
\author{Bartosz Such}
\affiliation{Center for Nanometer-Scale Science and Advanced Materials (NANOSAM), Faculty of Physics, Astronomy and Applied Computer Science, Jagiellonian University, ul. St.  Lojasiewicza 11, 30-348 Krakow, Poland}
\author{Marek Szymonski}
\affiliation{Center for Nanometer-Scale Science and Advanced Materials (NANOSAM), Faculty of Physics, Astronomy and Applied Computer Science, Jagiellonian University, ul. St.  Lojasiewicza 11, 30-348 Krakow, Poland}

\pacs{73.25.+i, 72.10.Bg, 72.80.Cw, 68.35.bg}
\keywords{surface conductance,  band bending, germanium, linear differential  equation}
\date{\today}
\begin{abstract}   
Surface conductance measurements on p-type doped germanium  show a small but systematic change to the surface conductivity  at different length scales. This effect is independent of the structure of the surface states. We interpret this phenomenon as a manifestation of conductivity changes beneath the surface. This hypothesis is  confirmed by an analysis of the classical current flow equation. We derive an integral formula for calculating of the effective surface conductivity as a function of the distance from a point source. Furthermore  we derive asymptotic values of the surface conductivity at small and large distances. The actual surface conductivity can only be sampled  close to the current source. At large distances, the conductivity measured on the surface corresponds to the bulk value.
\end{abstract}

\maketitle
\section{Introduction}
Surface conductance measurements at micron and sub-micron scales belong to popular experimental  routines characterizing semiconductor surfaces, see Ref.~\onlinecite{Hasegawa, Hofman1, Simons}. The motivation comes  from microelectronics~\cite{Simons} as well as from molecular electronics projects, and the results depend mainly on the system under investigation. The underlying physics may be determined mostly by surface states  \cite{Lublin,Wolkow}, by some bulk features or a mixture of these two limiting cases~\cite{Hoffmann, RSI}. In the first case, quantum transport theory  may be applied to understand the results, e.g. Ref.~\onlinecite{Lublin}. While in the latter case, classical conductance theory is effective. Surprisingly, the understanding of this well-established theory in the context of surface measurements is poor, making  even a qualitative interpretation of experimental data rather difficult. 

Conductance measurements on beveled surfaces have been used successfully  to reconstruct the doping profile below the surface with a spreading resistance analysis~\cite{spa1,spa2,spa3}. These methods, however, rely on refined numerical algorithms and offer no analytic insight. As such, they are an important tool in quantitative analysis but poorly contribute to our physical comprehension.

Recently, we have  reported surface conductance measurements obtained for germanium samples with an atomically clean (001) surface~\cite{APL}.   The results were interpreted within the classical theory without any reference to the surface states. We observed a slight but systematic change of the measured conductivity when varying  the distance between current sources. Our assumption was, that this was  due to the change of the conductivity beneath the surface as a result of Fermi level pinning, a phenomenon  resulting in variations of  charge carrier densities near the surface. Such modeling of the subsurface region  was numerically explored for the surface measurements on Si(111)\cite{Hoffmann}.  As we could not find any reference discussing  changes of the surface conductivity at different length scales, we decided to address the problem in more detail.  Surprisingly, general analytic conclusions can be made based on  a concise formula for the electrostatic potential profile due to the current point source. We show that a conductivity change in the bulk can not only make surface conductivity appear distance-dependent, but it can even vary the asymptotic behavior of the electrostatic potential. Furthermore,  the deviation from the bulk conductivity becomes experimentally inaccessible to surface measurements at large length scales.
 
Our paper is organized as follows. In sect.~\ref{Exp} we review the experimental data motivating our considerations. We reanalyze them in Appendix~\ref{AppB} in  light of our analytic developments. A brief discussion of the classical current flow equation is given in sect.~\ref{GenA}. Next, in sect.~\ref{GenB} we outline the so-called band bending phenomenon that motivates our model of the conductivity varying with the distance from the surface. In sect.~\ref{res} we outline our analytic findings while more rigorous treatment and technical details are discussed in Appendices~\ref{AppA} and~\ref{AppC} . We also demonstrate numerical results for two simple cases. Finally, in sect.~\ref{finitesize}, we comment on the possible finite size effects of the physical electrodes. We conclude with a summary and we point out  several interesting issues.

\section{Experimental data}\label{Exp}
We report surface conductivity obtained from surface conductance measurements for two p-type Ga-doped germanium samples cut from one wafer (MTI corporation, $\sigma=2-10 \ (\Omega cm)^{-1}$). Both the preparation and measurements were done  in an  ultra-high vacuum system~\cite{APL}. The data are obtained for a well-reconstructed  Ge(001) surface, a hydrogen terminated Ge(001) surface, and a partially dehydrogenated Ge(001):H surface. Ge samples were prepared using the procedure described in~Ref.~\onlinecite{APL}, where some data  for the bare surface have also been reported. The hydrogen termination of the surface was done following the procedure reported in Ref.~\onlinecite{Kolmer}. The partial dehydrogenation was achieved by irradiation of Ge(001):H with an electron beam from the Scanning Electron Microscope (SEM). As a result, in addition to an increased population of unsaturated dangling bonds a number of carbon atoms were adsorbed at the surface, as verified by  Scanning Auger Spectroscopy.

\begin{figure}
\resizebox{!}{0.3\textwidth}{
\begingroup
  \gdef\gplbacktext{}%
  \gdef\gplfronttext{}%
 \ifGPblacktext
    \def\colorrgb#1{}%
    \def\colorgray#1{}%
  \else
    \ifGPcolor
      \def\colorrgb#1{\color[rgb]{#1}}%
      \def\colorgray#1{\color[gray]{#1}}%
      \expandafter\def\csname LTw\endcsname{\color{white}}%
      \expandafter\def\csname LTb\endcsname{\color{black}}%
      \expandafter\def\csname LTa\endcsname{\color{black}}%
      \expandafter\def\csname LT0\endcsname{\color[rgb]{1,0,0}}%
      \expandafter\def\csname LT1\endcsname{\color[rgb]{0,1,0}}%
      \expandafter\def\csname LT2\endcsname{\color[rgb]{0,0,1}}%
      \expandafter\def\csname LT3\endcsname{\color[rgb]{1,0,1}}%
      \expandafter\def\csname LT4\endcsname{\color[rgb]{0,1,1}}%
      \expandafter\def\csname LT5\endcsname{\color[rgb]{1,1,0}}%
      \expandafter\def\csname LT6\endcsname{\color[rgb]{0,0,0}}%
      \expandafter\def\csname LT7\endcsname{\color[rgb]{1,0.3,0}}%
      \expandafter\def\csname LT8\endcsname{\color[rgb]{0.5,0.5,0.5}}%
    \else
      \def\colorrgb#1{\color{black}}%
      \def\colorgray#1{\color[gray]{#1}}%
      \expandafter\def\csname LTw\endcsname{\color{white}}%
      \expandafter\def\csname LTb\endcsname{\color{black}}%
      \expandafter\def\csname LTa\endcsname{\color{black}}%
      \expandafter\def\csname LT0\endcsname{\color{black}}%
      \expandafter\def\csname LT1\endcsname{\color{black}}%
      \expandafter\def\csname LT2\endcsname{\color{black}}%
      \expandafter\def\csname LT3\endcsname{\color{black}}%
      \expandafter\def\csname LT4\endcsname{\color{black}}%
      \expandafter\def\csname LT5\endcsname{\color{black}}%
      \expandafter\def\csname LT6\endcsname{\color{black}}%
      \expandafter\def\csname LT7\endcsname{\color{black}}%
      \expandafter\def\csname LT8\endcsname{\color{black}}%
    \fi
  \fi

  \setlength{\unitlength}{0.0500bp}%
  \begin{picture}(7200.00,5040.00)%
    \gplgaddtomacro\gplbacktext{%
      \csname LTb\endcsname%
      \put(682,704){\makebox(0,0)[r]{\strut{} \large{1}}}%
      \put(682,2061){\makebox(0,0)[r]{\strut{} \large{2}}}%
      \put(682,3418){\makebox(0,0)[r]{\strut{} \large{3}}}%
      \put(682,4775){\makebox(0,0)[r]{\strut{} \large{4}}}%
      \put(814,484){\makebox(0,0){\strut{}\large{ 0}}}%
      \put(2223,484){\makebox(0,0){\strut{}\large{ 4}}}%
      \put(3632,484){\makebox(0,0){\strut{} \large{8}}}%
      \put(5042,484){\makebox(0,0){\strut{}\large{ 12}}}%
      \put(6451,484){\makebox(0,0){\strut{} \large{16}}}%
      \put(176,2739){\rotatebox{-270}{\makebox(0,0){\large{$\sigma$ [($\Omega$cm)$^{-1}]$}}}}%
      \put(3808,154){\makebox(0,0){\large{$D$ [$\mu$m]}}}%
    }%
    \gplgaddtomacro\gplfronttext{%
      \csname LTb\endcsname%
      \put(5560,4602){\makebox(0,0)[r]{{Sample A -- Ge(001)}}}%
      \csname LTb\endcsname%
      \put(5790,4382){\makebox(0,0)[r]{{Sample A -- Ge(001):H}}}%
      \csname LTb\endcsname%
      \put(5630,4162){\makebox(0,0)[r]{{Sample A -- SEM irr.}}}%
      \csname LTb\endcsname%
      \put(5560,3942){\makebox(0,0)[r]{{Sample B -- Ge(001)}}}%
      \csname LTb\endcsname%
      \put(5790,3722){\makebox(0,0)[r]{{Sample B -- Ge(001):H}}}%
    }%
    \gplbacktext
    \put(0,0){\includegraphics{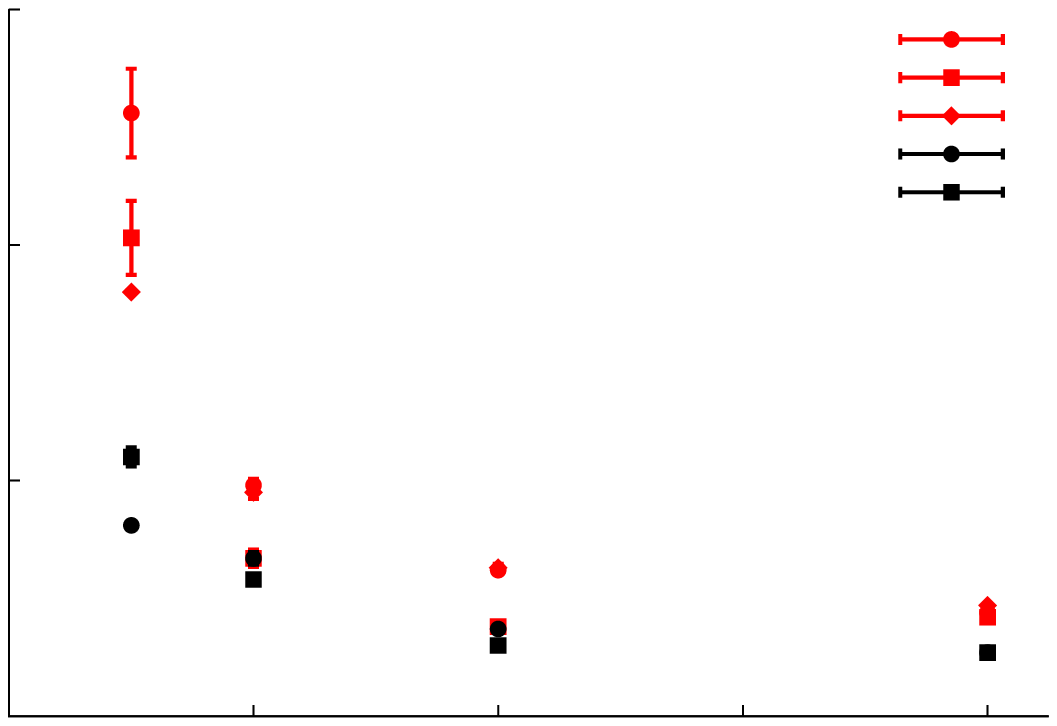}}%
    \gplfronttext
  \end{picture}%
\endgroup
}
\caption{ Conductivity $\sigma$ as a function of the distance $D$ between the current source and drain for three different germanium surfaces. The measurements were done using the four point-probe technique in a collinear arrangement outlined in details in Ref.~\cite{APL}. As some points coalesce, the data are also given in~Tab.~\ref{tab}. }
\end{figure}

The investigated surfaces have opposite transport properties due to the different electronic structures. There are well defined surface bands for the clean Ge(001) surface~\cite{ge001} close to the Fermi level.  The  Ge(001):H surface is isolating due to band positions far from the Fermi level~\cite{Kolmer,ARPES,Simons}. The  surface treated with the  electron beam is supposed to be a disordered system, hence it is weakly conducting at best. Nevertheless, the conductivity in all three cases is nearly the same. This is an interesting property of the Ge samples that we shall address in a future publication. For our considerations here, it is important that the surface contribution  to the current flow can be disregarded. As such, we can use the classical description of the current  suited for transport phenomena in bulk. \\
Surface conductance measurements were done using a four point probe technique as outlined in Ref.~\onlinecite{APL}. For a given distance $D$ between the  current source and drain we measured the voltage drop between two additional probes. All four electrodes were collinear with the  inner electrodes probing the voltage drop. The geometry was symmetric with respect to the point equidistant to the electrodes supplementing the current. For every $D$ the voltage drop was measured for several positions of the inner electrodes. This allowed us to calculate  the conductivity  for every $D$ separately using a simple fitting procedure. The data were consistent with the model for  three-dimensional current flow. For n-type doped samples two dimensional currents were observed, which we do not address here neither experimentally nor theoretically.\\
The data shown in Fig. 1 and Tab.~\ref{tab} were obtained from two samples   with different tips supplementing the current.  What is of interest here is the  fact that the  conductivity resulting from these measurements changes with the distance $D$.  These changes seem to follow a deterministic pattern rather than being of a stochastic origin. This led us to  believe that these changes are not an artifact  but a physical phenomenon. Below, we shall put this statement on  firm ground. We revisit the experimental data in Appendix~\ref{AppB}. 

\begingroup
\squeezetable
 \begin{table}[b]
 \begin{tabular}{c||c|c|c||c|c}
 D&\multicolumn{3}{c||}{Sample A}& \multicolumn{2}{c}{Sample B}\\
 \cline{2-6}
 [$\mu m$] & Ge(001) & Ge(001):H & SEM irr. & Ge(001) & Ge(001):H\\
\hline \hline
2  & 3.56 $\pm$ 0.19 & 3.03 $\pm$ 0.16 & 2.80 $\pm$ 0.06 & 1.81 $\pm$ 0.06 & 2.10 $\pm$ 0.04 \\
 \hline
4 & 1.98 $\pm$ 0.03 & 1.67 $\pm$ 0.04 & 1.95 $\pm$ 0.03 &  1.67 $\pm$ 0.03 & 1.58 $\pm$ 0.01 \\
 \hline
8  & 1.62 $\pm$ 0.02 & 1.38 $\pm$ 0.01 & 1.63 $\pm$ 0.02 & 1.37 $\pm$ 0.01 & 1.30 $\pm$ 0.01 \\
 \hline
16 & 1.44 $\pm$ 0.01 & 1.42 $\pm$ 0.01 & 1.47 $\pm$ 0.01 & 1.27 $\pm$ 0.01 & 1.27 $\pm$ 0.01\\
 \hline
\end{tabular}
\caption{Experimental values of conductivity (in $(\Omega$cm$)^{-1}$) for various surfaces and distances shown in Fig. 1.  SEM irr. denotes the surface irradiated by the electron beam from SEM. \label{tab}}
\end{table} 
\endgroup

\section{General framework} \label{intro}
\subsection{Classical model of the current flow}\label{GenA}
The classical equation governing the current flow follows from the Maxwell equations~\cite{Landau}. It can be derived in three steps. First, as a stationary solution is addressed, all time derivatives in the Maxwell equations are set to zero. In turn, the electric field $\mathbf{E}$ is a curl-free field
\begin{displaymath}
\nabla \times \mathbf{E}=0,
\end{displaymath}
giving rise to the notion of the electrostatic potential $\Phi(\mathbf{x})$, such that electric field is given by its gradient, $\mathbf{E}=\nabla\Phi$. Second, a  phenomenological input to the theory is needed to relate the electrostatic field and the current density $\textbf{j}(\textbf{x})$, where a linear relation is usually assumed
\begin{equation}
\textbf{j}(\textbf{x})=\sigma(\textbf{x})\nabla \Phi (\textbf{x}),
\end{equation}
where  the conductivity $\sigma$ is a scalar.   The coefficient $\sigma$ is a material property, independent of measurement or the sample  geometry. Third, from the continuity equation ($\nabla \cdot \mathbf{j}=0$) we obtain the relation
\begin{equation}\label{current1}
\nabla\left(\sigma\nabla\Phi\right)=0.
\end{equation}
This equation is valid in regions where there are no current sources or drains, which can be taken into account by modifying the right-hand side of the above equation~\cite{Hoffmann,functMat}. 

\subsection{Conductivity}\label{GenB}
The proportionality coefficient $\sigma$ between the current density and the electrostatic field has its microscopic interpretation.  For semiconductors it is expressed in terms of the material properties,
\begin{equation}
\sigma= n_{e}v_{e}+n_{h}v_{h},\label{sigma}
\end{equation}
where $n_{e/h}$ stands for the electron/hole density and $v_{e/h}$ is their mobility, respectively; see Ref.~\onlinecite{sze}. 

Band bending is a phenomenon that calls for considering models with the conductivity \mbox{$\sigma=\sigma(z)$} changing with the distance $z\geq 0$ from the surface (at $z = 0$).  It is explained in many textbooks, e.g. Ref.~\onlinecite{Monch}, so we confine ourselves to a succinct resume in terms of a self-consistent calculation scheme, and we illustrate the main facts about it  in Fig.~2.  
The phenomenon is due to  surface states whose energy is determined in relation to bands, they may appear below the valence band, above the conductance band, or within the band gap. Assume there is no shift of the bands against the Fermi level at the surface. As such, the surface states accumulate a large electric charge. This  gives rise to the creation of an electrostatic field that may be seen in  the semiclassical (envelope) approximation of a shift of the band positions with respect to the Fermi level. But such shifts change the electric charge at the surface.  In real crystals, we observe  the surface Fermi level at the position arriving as a self-consistent solution of the problem sketched above. 
Thus, in equilibrium the semiconductor surface is slightly charged and is accompanied by an electrostatic field. Due to this field, the relative position of bands with respect to the Fermi level varies near the surface, and in this region the  number of current carriers may be different from their bulk values. As indicated by eq.~(\ref{sigma}), a variation of the conductivity follows naturally. A similar model can also be found in Ref.~\onlinecite{Hoffmann}. More quantitative considerations allowing for band bending and carrier density  profiles  can be found in Ref.~\onlinecite{King,Multiscale}. We note, that treating  band bending and current flow as independent phenomena is not entirely correct as both of these phenomena refer to the electrostatic potential.   However, at sufficiently large distances from the current source, the current-associated  field is so weak that it cannot substantially change the band bending. It is also not certain, if the above model is straightforwardly applicable to germanium. However,  strong Fermi level pinning in germanium at the valence band is beyond any doubt~\cite{APL}.
\begin{figure}
\includegraphics[scale=0.6]{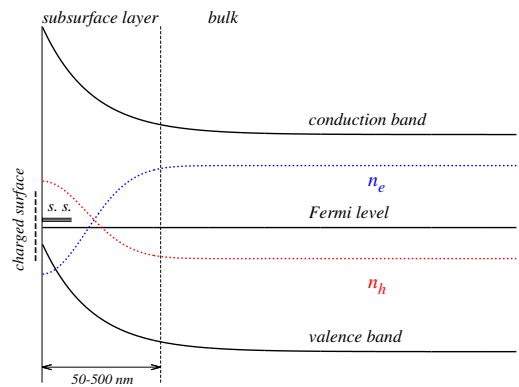}
\caption{ A schematic picture explaining the band bending and associated change in the number of electrons and holes. The acronym s.~s. stands for surface states, $n_{e}$ and $n_{h}$ for the electron and hole density, respectively.  The width of the subsurface region depends on band bending and bulk doping, usually it is in the range given in the picture. The   depicted situation with upwards band bending corresponds to germanium, where the valence band appears very close to the Fermi level. }
\end{figure} 

\section{Surface current from point source}\label{res}
\subsection{Transformation of the current equation}
Now, we turn to eq.~(\ref{current1}).  We are interested in solving the equation in semi-space, i.e., with unrestricted two Cartesian coordinates $x$ and $y$ and the third one constrained to the upper semi-axis $ z\ge 0$.
As already mentioned, to take into account current sources and drains we need to modify eq.~(\ref{current1}) by adding a source term. A common choice is the point-source.  As such, we concentrate on the following equation,
\begin{equation}\label{source1}
\nabla\left(\sigma\nabla\Phi\right)=\delta(\textbf{x}-\textbf{x}').
\end{equation}
To fully formulate the model, we state the boundary condition 
\begin{equation}\label{boundary1}
\left.\frac{\partial \Phi}{\partial z}\right|_{z=0}=0,
 \end{equation}
which ensures that there is no current flowing through the plane  $z=0$.
Eq.~(\ref{source1}) is a linear equation and may be transformed to a familiar form of a Schr\"odinger operator if we switch from the function $\Phi$ to a function~$\xi$ defined by the relation
\begin{equation}\label{transf}
\Phi=\frac{\xi}{\sqrt{\sigma}}.
\end{equation}
The resulting form of  eq.~(\ref{source1}) yields
\begin{equation}\label{defL}
\sqrt{\sigma} \underbrace{ \left[-\Delta + \frac{\Delta\sqrt{\sigma}}{\sqrt{\sigma}}\right]}_{\hat{L}}\xi=\delta(\textbf{x}-\textbf{x}').
\end{equation}
Further, in the case of interest $\sigma$ depends only on the $z$ coordinate. We shall use the representation of the Dirac delta in terms of eigenfunctions and eignenvalues of the operator $\hat{L}$. The eigenproblem factorizes,  and in two dimensions ($x$ and $y$) it is trivial. All complexity is in the one-dimensional equation
\begin{equation}\label{Schrod1}
 \left[-\frac{d^{2}}{dz^{2}} + \frac{\frac{d^{2}}{dz^{2}}\sqrt{\sigma(z)}}{\sqrt{\sigma(z)}}\right]\psi(k;z)=k^{2}\psi(k;z).
\end{equation}
Now we study this equation, as both its eigenfunctions and eigenvalues appear in the solution of eq.~(\ref{source1}) and~(\ref{defL}). In analogy to the Schr{\"o}dinger equation, we shall refer to the term 
\begin{equation}
V(z)=\frac{\frac{d^{2}}{dz^{2}}\sqrt{\sigma(z)}}{\sqrt{\sigma(z)}}
\end{equation}
as the potential. If the potential vanishes, the corresponding eq.~(\ref{Schrod1}) is called potential-free. Noteworthy, the  correspondence between the current equation and the time-independent  Schr{\"o}dinger equation is a formal feature, which will facilitate our reasoning. However, one should bear in mind, that the current flow equation we deal with is a classical theory of the electrostatic field, as shown in sect.~\ref{GenA}.\\
Transformation~(\ref{transf}) mixes the electrostatic potential $\Phi$ and the function $\sigma$. The boundary condition~(\ref{boundary1}) then reads 
\begin{eqnarray}
\partial_{z}\psi(k;0)=0 & \mbox{ \ \ \ for \ \ \ } &\partial_{z}\sigma(0)=0, \label{bound2}\\
\frac{\partial_{z} \psi(k;0)}{\psi(k;0)}=\frac{\partial_{z} \sigma(0)}{2\sigma(0)}  & \mbox{  \ \ \ for \ \ \  } & \partial_{z} \sigma(0) \neq 0,\label{bound3}
\end{eqnarray}
where $\partial_{z}\psi(k;0)$  is  $d \psi(k,z)/dz $ taken at $z=0$. 

\subsection{The potential}\label{potential}
Eq.~(\ref{Schrod1}) binds the potential  and the conductivity $\sigma$ in a highly non-trivial way. To shed some light on its physical meaning we introduce a function $\eta$  
\begin{equation}
\eta(z)=\frac{1}{2} \ln{\frac{\sigma(z)}{\sigma(0)}}.
\end{equation} 
The function reflects how the conductivity, and hence the carrier density as implied by eq.~(\ref{sigma}), changes on the  logarithmic scale. The potential can be rewritten in the following form
\begin{equation}
V(z)=\left(\partial_{z} \eta \right)^{2}+\partial_{z}^{2}\eta.
\end{equation}
We observe, that the potential is  independent of the absolute value of $\sigma$, i.e. multiplication of $\sigma$ with any number does not alter the potential.  As the potential contains   terms  with the first and second derivative of $\sigma$, the range of the change is important. A given potential value may be  obtained for a small but abrupt change of $\sigma$ as well as for a large deviation spread on a long distance. If the variation  of  $\sigma$  is slow enough, the resulting potential is negligible. We assume it is the case for large $z$ as the bulk conductivity is well defined. The presence of the term $\partial_{z}^{2}\eta$ appears to be slightly inconvenient from a physical standpoint. It requires   accurate sampling of $\sigma$ if it is to define a physical model.  In this paper, we are interested in potentials that could be termed as  small perturbations around the zero valued potential. 

Fig.~3 and~4 show two potentials corresponding to two conductance profiles illustrating the formation of  accumulation and depletion layers near the surface. As will be shown, physically relevant results are obtained if condition~(\ref{bound2}) is assumed. This is, however, at odds with the classical~\cite{Monch} and quantum~\cite{King} modeling of the carrier density.  As such, these conductivity profiles do not appear as results of a theoretical calculation.  A profile of $\sigma(z)$ approximately reproducing experimental data is given in Appendix~\ref{AppB}. 

\begin{figure}
\resizebox{!}{0.35\textwidth}{
\begingroup
  \gdef\gplbacktext{}%
  \gdef\gplfronttext{}%
 \ifGPblacktext
    \def\colorrgb#1{}%
    \def\colorgray#1{}%
  \else
    \ifGPcolor
      \def\colorrgb#1{\color[rgb]{#1}}%
      \def\colorgray#1{\color[gray]{#1}}%
      \expandafter\def\csname LTw\endcsname{\color{white}}%
      \expandafter\def\csname LTb\endcsname{\color{black}}%
      \expandafter\def\csname LTa\endcsname{\color{black}}%
      \expandafter\def\csname LT0\endcsname{\color[rgb]{1,0,0}}%
      \expandafter\def\csname LT1\endcsname{\color[rgb]{0,1,0}}%
      \expandafter\def\csname LT2\endcsname{\color[rgb]{0,0,1}}%
      \expandafter\def\csname LT3\endcsname{\color[rgb]{1,0,1}}%
      \expandafter\def\csname LT4\endcsname{\color[rgb]{0,1,1}}%
      \expandafter\def\csname LT5\endcsname{\color[rgb]{1,1,0}}%
      \expandafter\def\csname LT6\endcsname{\color[rgb]{0,0,0}}%
      \expandafter\def\csname LT7\endcsname{\color[rgb]{1,0.3,0}}%
      \expandafter\def\csname LT8\endcsname{\color[rgb]{0.5,0.5,0.5}}%
    \else
      \def\colorrgb#1{\color{black}}%
      \def\colorgray#1{\color[gray]{#1}}%
      \expandafter\def\csname LTw\endcsname{\color{white}}%
      \expandafter\def\csname LTb\endcsname{\color{black}}%
      \expandafter\def\csname LTa\endcsname{\color{black}}%
      \expandafter\def\csname LT0\endcsname{\color{black}}%
      \expandafter\def\csname LT1\endcsname{\color{black}}%
      \expandafter\def\csname LT2\endcsname{\color{black}}%
      \expandafter\def\csname LT3\endcsname{\color{black}}%
      \expandafter\def\csname LT4\endcsname{\color{black}}%
      \expandafter\def\csname LT5\endcsname{\color{black}}%
      \expandafter\def\csname LT6\endcsname{\color{black}}%
      \expandafter\def\csname LT7\endcsname{\color{black}}%
      \expandafter\def\csname LT8\endcsname{\color{black}}%
    \fi
  \fi
  \setlength{\unitlength}{0.0500bp}%
  \begin{picture}(7200.00,5040.00)%
    \gplgaddtomacro\gplbacktext{%
      \csname LTb\endcsname%
      \put(1210,704){\makebox(0,0)[r]{\strut{}\large{$-4\cdot 10^{-5}$}}}%
      \put(1210,1722){\makebox(0,0)[r]{\strut{}\large{$-2\cdot 10^{-5}$}}}%
      \put(1210,2739){\makebox(0,0)[r]{\strut{} \large{0}}}%
      \put(1210,3757){\makebox(0,0)[r]{\strut{}\large{ $2\cdot 10^{-5}$}}}%
      \put(1210,4775){\makebox(0,0)[r]{\strut{} \large{$4\cdot 10^{-5}$}}}%
      \put(1342,484){\makebox(0,0){\strut{}\large{ 0}}}%
      \put(2825,484){\makebox(0,0){\strut{}\large{ 100}}}%
      \put(4308,484){\makebox(0,0){\strut{}\large{ 200}}}%
      \put(5791,484){\makebox(0,0){\strut{}\large{ 300}}}%
      \put(5923,1043){\makebox(0,0)[l]{\strut{}\large{ 0.83}}}%
      \put(5923,4436){\makebox(0,0)[l]{\strut{} \large{1.00}}}%
      \put(176,2739){\rotatebox{-270}{\makebox(0,0){\large{$V(z) [nm^{-2}]$}}}}%
      \put(6500,2739){\makebox(0,0){\color{red} \LARGE{$\frac{\sigma(z)}{\sigma(0)}$}}}%
      \put(3566,154){\makebox(0,0){\large{$z \ [nm]$}}}%
    }%
    \gplgaddtomacro\gplfronttext{%
    }%
    \gplbacktext
    \put(0,0){\includegraphics{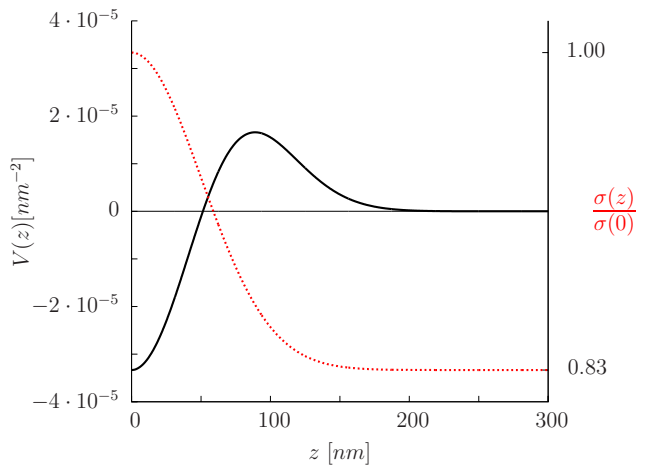}}%
    \gplfronttext
  \end{picture}%
\endgroup
}
\caption{ Gaussian conductance profile describing a tiny enhancement of the conductance at the surface (red dotted line) and related potential $V$ (black solid line). The conductance profile reads   $\sigma(z) = \sigma(0) \left[ 0.833 + 0.267\exp(-0.0002 z^{2}) \right]$. \label{pot1}  }
\end{figure}
\begin{figure}
\resizebox{!}{0.35\textwidth}{
\begingroup
  \gdef\gplbacktext{}%
  \gdef\gplfronttext{}%
 \ifGPblacktext
    \def\colorrgb#1{}%
    \def\colorgray#1{}%
  \else
    \ifGPcolor
      \def\colorrgb#1{\color[rgb]{#1}}%
      \def\colorgray#1{\color[gray]{#1}}%
      \expandafter\def\csname LTw\endcsname{\color{white}}%
      \expandafter\def\csname LTb\endcsname{\color{black}}%
      \expandafter\def\csname LTa\endcsname{\color{black}}%
      \expandafter\def\csname LT0\endcsname{\color[rgb]{1,0,0}}%
      \expandafter\def\csname LT1\endcsname{\color[rgb]{0,1,0}}%
      \expandafter\def\csname LT2\endcsname{\color[rgb]{0,0,1}}%
      \expandafter\def\csname LT3\endcsname{\color[rgb]{1,0,1}}%
      \expandafter\def\csname LT4\endcsname{\color[rgb]{0,1,1}}%
      \expandafter\def\csname LT5\endcsname{\color[rgb]{1,1,0}}%
      \expandafter\def\csname LT6\endcsname{\color[rgb]{0,0,0}}%
      \expandafter\def\csname LT7\endcsname{\color[rgb]{1,0.3,0}}%
      \expandafter\def\csname LT8\endcsname{\color[rgb]{0.5,0.5,0.5}}%
    \else
      \def\colorrgb#1{\color{black}}%
      \def\colorgray#1{\color[gray]{#1}}%
      \expandafter\def\csname LTw\endcsname{\color{white}}%
      \expandafter\def\csname LTb\endcsname{\color{black}}%
      \expandafter\def\csname LTa\endcsname{\color{black}}%
      \expandafter\def\csname LT0\endcsname{\color{black}}%
      \expandafter\def\csname LT1\endcsname{\color{black}}%
      \expandafter\def\csname LT2\endcsname{\color{black}}%
      \expandafter\def\csname LT3\endcsname{\color{black}}%
      \expandafter\def\csname LT4\endcsname{\color{black}}%
      \expandafter\def\csname LT5\endcsname{\color{black}}%
      \expandafter\def\csname LT6\endcsname{\color{black}}%
      \expandafter\def\csname LT7\endcsname{\color{black}}%
      \expandafter\def\csname LT8\endcsname{\color{black}}%
    \fi
  \fi
  \setlength{\unitlength}{0.0500bp}%
  \begin{picture}(7200.00,5040.00)%
    \gplgaddtomacro\gplbacktext{%
      \csname LTb\endcsname%
      \put(1474,704){\makebox(0,0)[r]{\strut{}\large{$-3\cdot 10^{-5}$}}}%
      \put(1474,2061){\makebox(0,0)[r]{\strut{} \large{0}}}%
      \put(1474,3418){\makebox(0,0)[r]{\strut{}\large{ $3\cdot 10^{-5}$}}}%
      \put(1474,4775){\makebox(0,0)[r]{\strut{} \large{$6\cdot 10^{-5}$}}}%
      \put(1606,484){\makebox(0,0){\strut{}\large{ 0}}}%
      \put(3227,484){\makebox(0,0){\strut{} \large{100}}}%
      \put(4848,484){\makebox(0,0){\strut{} \large{200}}}%
      \put(5791,1286){\makebox(0,0)[l]{\strut{}\large{ 1}}}%
      \put(5791,4193){\makebox(0,0)[l]{\strut{} \large{1.25}}}%
      \put(176,2739){\rotatebox{-270}{\makebox(0,0){\large{$V(z) \ [nm^{-2}]$}}}}%
      \put(6400,2739){\makebox(0,0){\color{red}\LARGE{$\frac{\sigma(z)}{\sigma(0)}$}}}%
      \put(3632,154){\makebox(0,0){\large{$z [nm]$}}}%
    }%
    \gplgaddtomacro\gplfronttext{%
    }%
    \gplbacktext
    \put(0,0){\includegraphics{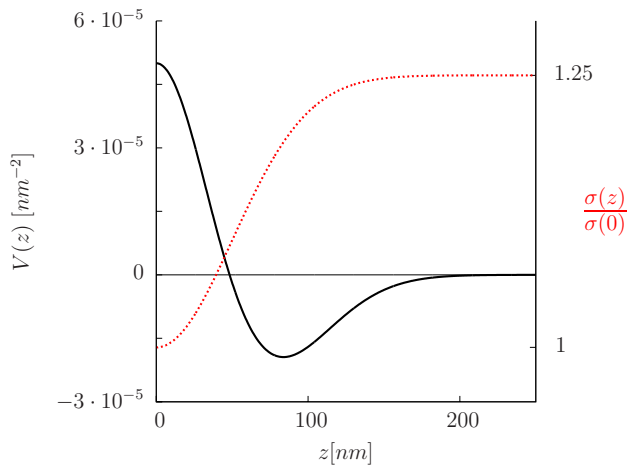}}%
    \gplfronttext
  \end{picture}%
\endgroup
}
\caption{ Gaussian conductance profile describing a small deficiency of the conductance at the surface (red dotted line) and related potential $V$ (black solid line). The conductance profile reads   $\sigma(z) = \sigma(0) \left[ 1.25 - 0.25\exp(-0.0002 z^{2}) \right]$.  \label{pot2}}
\end{figure}

\subsection{Electrostatic potential at the surface}
Not much can be said in general about solutions $\Phi$ of eq.~(\ref{source1}). In the potential-free case, the Green function may be calculated exactly, as shown in the Appendix~\ref{AppC}. However, we can infer a lot about the  surface profile of the Green function,
\begin{displaymath}
\phi(r)=\Phi(x,y,z=0;x',y',z'=0),
\end{displaymath}
where radial coordinate $r=\sqrt{(x-x')^{2}+(y-y')^{2}}$ is introduced and the point source is located  at $(x',y',z'=0)$.
As shown in  Appendix~\ref{AppA},  $\phi(r)$  satisfies the following integral formula
\begin{equation}\label{integral1}
\dfrac{\sigma(0)}{2 \pi}\phi(r)=\frac{1}{r}\int_{0}^{\infty} du \ K_{0}(u)\psi^{2}\left(\frac{u}{r},0\right),
\end{equation}
where $K_{0}(u)$ is the modified Bessel function of the second kind  of  zeroth order.  This function  explodes logarithmically at the origin and vanishes exponentially for large values of $u$; see Fig.~5. For the assumed potentials, the function $\psi(k;0)$ is bounded and does not quickly vary. As such, the asymptotic behaviors of $\phi(r)$ may be deduced from the above formula. Here we summarize the results elaborated in detail in  Appendix~\ref{AppA} and informally explained in Fig. 5.   At this point we do not care much about the multiplicative prefactor of $\phi(r)$. In sect.~\ref{eff}, we will set its normalization to suit  the physical context.   First, we report the short-range limit ($r\to 0$), for which we have the following relation:
\begin{equation}\label{shortrange}
\dfrac{\sigma(0)}{2 \pi}\phi(r)\underset{r\to 0} {\longrightarrow}\frac{1}{r}\int_{0}^{\infty} du \ K_{0}(u)\psi^{2}(\infty;0)=\frac{\pi}{2 r}.
\end{equation}
The same result would be obtained if there were a uniform conductivity $\sigma(0)$ in the whole sample. This is reasonable, as close to the source the current explores only  a thin layer in the surface neighborhood where the conductivity can be considered constant. It is the behavior of $\psi(k;0)$ for large $k$ that is important here.

There are two  asymptotic relations for large $r$ depending on whether condition~(\ref{bound2}) or~(\ref{bound3}) is considered. For~(\ref{bound2}) we can assume, that $\psi(0;0)\neq 0$, and then
\begin{equation}\label{longrange}
\dfrac{\sigma(0)}{2 \pi}\phi(r)\underset{r\to\infty} {\longrightarrow}\frac{1}{r}\int_{0}^{\infty} du \ K_{0}(u)\psi^{2}(0;0)=\frac{\pi \frac{\sigma(0)}{\sigma(\infty)}}{2 r},
\end{equation}  
where $\sigma(\infty)$ corresponds to the bulk value of the conductivity, $\sigma(\infty)=\lim_{z\to\infty}\sigma(z)$. It is a very intriguing fact, that the asymptotic values  of conductivity are given by its surface and bulk values, no matter how complicated the conductivity profile beneath the surface is.

The second boundary condition~(\ref{bound3}) gives rise to a modified behavior at large distances $\phi(r)\sim r^{-3}$  if $\partial_{z}\sigma(0)>0$. The case of negative $\partial_{z}\sigma(0)$ is not treated here for the bound state it has in the spectrum   (for more details see  Appendix~\ref{AppA}). We do not know any experimental results demonstrating such a quick voltage drop on the surface. This is why it appears to be   a hint as to how the conductivity should be modeled close to surfaces. Noteworthy, numerical algorithms often deal with thin slices of constant conductivity~\cite{spa2,voigtlander} parallel to the surface. As such, they assume condition~(\ref{boundary1}).    \\
\begin{figure}
\resizebox{!}{0.325\textwidth}{
\begingroup
  \gdef\gplbacktext{}%
  \gdef\gplfronttext{}%
 \ifGPblacktext
    \def\colorrgb#1{}%
    \def\colorgray#1{}%
  \else
    \ifGPcolor
      \def\colorrgb#1{\color[rgb]{#1}}%
      \def\colorgray#1{\color[gray]{#1}}%
      \expandafter\def\csname LTw\endcsname{\color{white}}%
      \expandafter\def\csname LTb\endcsname{\color{black}}%
      \expandafter\def\csname LTa\endcsname{\color{black}}%
      \expandafter\def\csname LT0\endcsname{\color[rgb]{1,0,0}}%
      \expandafter\def\csname LT1\endcsname{\color[rgb]{0,1,0}}%
      \expandafter\def\csname LT2\endcsname{\color[rgb]{0,0,1}}%
      \expandafter\def\csname LT3\endcsname{\color[rgb]{1,0,1}}%
      \expandafter\def\csname LT4\endcsname{\color[rgb]{0,1,1}}%
      \expandafter\def\csname LT5\endcsname{\color[rgb]{1,1,0}}%
      \expandafter\def\csname LT6\endcsname{\color[rgb]{0,0,0}}%
      \expandafter\def\csname LT7\endcsname{\color[rgb]{1,0.3,0}}%
      \expandafter\def\csname LT8\endcsname{\color[rgb]{0.5,0.5,0.5}}%
    \else
      \def\colorrgb#1{\color{black}}%
      \def\colorgray#1{\color[gray]{#1}}%
      \expandafter\def\csname LTw\endcsname{\color{white}}%
      \expandafter\def\csname LTb\endcsname{\color{black}}%
      \expandafter\def\csname LTa\endcsname{\color{black}}%
      \expandafter\def\csname LT0\endcsname{\color{black}}%
      \expandafter\def\csname LT1\endcsname{\color{black}}%
      \expandafter\def\csname LT2\endcsname{\color{black}}%
      \expandafter\def\csname LT3\endcsname{\color{black}}%
      \expandafter\def\csname LT4\endcsname{\color{black}}%
      \expandafter\def\csname LT5\endcsname{\color{black}}%
      \expandafter\def\csname LT6\endcsname{\color{black}}%
      \expandafter\def\csname LT7\endcsname{\color{black}}%
      \expandafter\def\csname LT8\endcsname{\color{black}}%
    \fi
  \fi
  \setlength{\unitlength}{0.0500bp}%
  \begin{picture}(7200.00,5040.00)%
    \gplgaddtomacro\gplbacktext{%
      \csname LTb\endcsname%
      \put(726,704){\makebox(0,0)[r]{\strut{} \large{$0$}}}%
      \put(726,1518){\makebox(0,0)[r]{\strut{} \large{$0.5$}}}%
      \put(726,2332){\makebox(0,0)[r]{\strut{} \large{$1$}}}%
      \put(726,3147){\makebox(0,0)[r]{\strut{} \large{$1.5$}}}%
      \put(726,3961){\makebox(0,0)[r]{\strut{} \large{$2$}}}%
      \put(726,4775){\makebox(0,0)[r]{\strut{} \large{$2.5$}}}%
      \put(858,484){\makebox(0,0){\strut{} \large{$0$}}}%
      \put(2047,484){\makebox(0,0){\strut{} \large{$1$}}}%
      \put(3236,484){\makebox(0,0){\strut{} \large{$2$}}}%
      \put(4425,484){\makebox(0,0){\strut{} \large{$3$}}}%
      \put(5614,484){\makebox(0,0){\strut{} \large{$4$}}}%
      \put(6803,484){\makebox(0,0){\strut{} \large{$5$}}}%
      \put(3830,154){\makebox(0,0){\Large{$u$}}}%
      \colorrgb{0.00,0.00,0.00}%
      \put(1215,3798){\makebox(0,0)[l]{\large{$\sim -\ln(u)$}}}%
      \put(4920,1100){\makebox(0,0)[l]{\large{$\sim \dfrac{\exp(-u)}{\sqrt{u}}$}}}%
      \put(2998,1160){\makebox(0,0)[l]{\large{$K_{0}(u)$}}}%
      \colorrgb{0.00,0.00,1.00}%
      \put(5614,2658){\makebox(0,0)[l]{\color{blue} { $\psi^{2}\left(\dfrac{u}{12};0\right)$}}}%
      \colorrgb{1.00,0.00,0.00}%
      \put(3400,2050){\makebox(0,0){\color{red} {$\psi^{2}\left(\dfrac{u}{0.6};0\right)$}}}%
    }%
    \gplgaddtomacro\gplfronttext{%
    }%
    \gplbacktext
    \put(0,0){\includegraphics{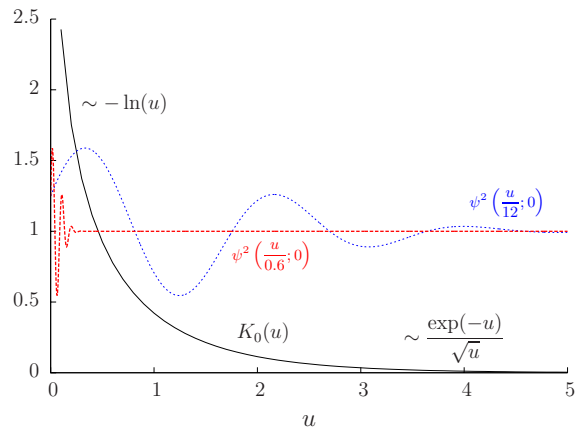}}%
    \gplfronttext
  \end{picture}%
\endgroup
}
\caption{ The black solid curve corresponds to  the function $K_{0}(u)$  with a logarithmic divergence close to $u=0$ and vanishing exponentially for $u\to \infty$. Two other curves  correspond to  some function $\psi^{2}(u/r;0)$ with $r=0.6$ (red and dashed) and $r=12$ (blue dotted). Due to the quick decrease of $K_{0}(u)$,  only the region $(0;3.5)$ contributes significantly to the integral, eq.~(\ref{integral1}). For small values of~$r$, the function $\psi^{2}$ gets compressed and it has the limiting value 1 nearly on the whole interval $(0,3.5)$. For large values of $r$, the function $\psi^{2}$ is stretched, so that for sufficiently large~$r$ it is nearly constant (equal to $\psi(0;0)$) on the interval. The compression and stretching is responsible for the obtained asymptotic behavior.}
\end{figure}

\subsection{Effective surface conductivity}\label{eff}
The asymptotic description obtained for the condition~(\ref{bound2}) motivates introducing a function $\sigma_{eff}(r)$ which is defined as
\begin{equation}
\frac{\sigma(0)}{\sigma_{eff}(r)}=\frac{2}{\pi} \int_{0}^{\infty} du \ 
K_{0}(u)\psi^{2}\left(\frac{u}{r},0\right).
\end{equation}
To justify this definition, we  write  the electrostatic potential $\Phi$ in an infinitesimal range around  the source,
\begin{equation}\label{norm}
\Phi(r_{1},z)=\frac{I}{2 \pi \sigma(0)}\frac{1}{\sqrt{r_{1}^{2}+z^{2}}},
\end{equation}
where $I$ is the current supplemented. In this way we get a normalization factor for the whole range of $r$.
As a consequence of eq.~(\ref{integral1}) one can write the following for any~$r$ and~$r_{1}$
\begin{equation}
\phi(r)=\frac{r^{-1}\int_{0}^{\infty} du \ K_{0}(u)\psi^{2}\left(\frac{u}{r},0\right)}{r_{1}^{-1}\int_{0}^{\infty} du \ K_{0}(u)\psi^{2}\left(\frac{u}{r_{1}},0\right)} \phi(r_{1}).
\end{equation}
For small $r_{1}$ both eq.~(\ref{shortrange}) and~(\ref{norm}) hold. They may be  plugged into the above formula resulting in the relation
\begin{equation}
\phi(r)=\frac{I}{2 \pi \sigma_{eff}(r)}\frac{1}{r }.
\end{equation}   
Thus, $\sigma_{eff}(r)$ may be viewed as an effective surface conductivity. It reflects the fact that  the measured surface conductivity varies with the distance from the source due to changes in the conductivity below.

Fig.~6 and~7 show the effective conductivity obtained for the potentials shown in Fig.~3 and~4. The results come from numerical integration of eq.~(\ref{integral1}) with numerically calculated functions $\psi(k;z)$.
\begin{figure}
\resizebox{!}{0.30\textwidth}{
\begingroup
  \gdef\gplbacktext{}%
  \gdef\gplfronttext{}%
 \ifGPblacktext
    \def\colorrgb#1{}%
    \def\colorgray#1{}%
  \else
    \ifGPcolor
      \def\colorrgb#1{\color[rgb]{#1}}%
      \def\colorgray#1{\color[gray]{#1}}%
      \expandafter\def\csname LTw\endcsname{\color{white}}%
      \expandafter\def\csname LTb\endcsname{\color{black}}%
      \expandafter\def\csname LTa\endcsname{\color{black}}%
      \expandafter\def\csname LT0\endcsname{\color[rgb]{1,0,0}}%
      \expandafter\def\csname LT1\endcsname{\color[rgb]{0,1,0}}%
      \expandafter\def\csname LT2\endcsname{\color[rgb]{0,0,1}}%
      \expandafter\def\csname LT3\endcsname{\color[rgb]{1,0,1}}%
      \expandafter\def\csname LT4\endcsname{\color[rgb]{0,1,1}}%
      \expandafter\def\csname LT5\endcsname{\color[rgb]{1,1,0}}%
      \expandafter\def\csname LT6\endcsname{\color[rgb]{0,0,0}}%
      \expandafter\def\csname LT7\endcsname{\color[rgb]{1,0.3,0}}%
      \expandafter\def\csname LT8\endcsname{\color[rgb]{0.5,0.5,0.5}}%
    \else
      \def\colorrgb#1{\color{black}}%
      \def\colorgray#1{\color[gray]{#1}}%
      \expandafter\def\csname LTw\endcsname{\color{white}}%
      \expandafter\def\csname LTb\endcsname{\color{black}}%
      \expandafter\def\csname LTa\endcsname{\color{black}}%
      \expandafter\def\csname LT0\endcsname{\color{black}}%
      \expandafter\def\csname LT1\endcsname{\color{black}}%
      \expandafter\def\csname LT2\endcsname{\color{black}}%
      \expandafter\def\csname LT3\endcsname{\color{black}}%
      \expandafter\def\csname LT4\endcsname{\color{black}}%
      \expandafter\def\csname LT5\endcsname{\color{black}}%
      \expandafter\def\csname LT6\endcsname{\color{black}}%
      \expandafter\def\csname LT7\endcsname{\color{black}}%
      \expandafter\def\csname LT8\endcsname{\color{black}}%
    \fi
  \fi
  \setlength{\unitlength}{0.0500bp}%
  \begin{picture}(7200.00,5040.00)%
    \gplgaddtomacro\gplbacktext{%
      \csname LTb\endcsname%
      \put(1078,913){\makebox(0,0)[r]{\large{ 0.84}}}%
      \put(1078,1748){\makebox(0,0)[r]{\large{ 0.88}}}%
      \put(1078,2583){\makebox(0,0)[r]{\large{ 0.92}}}%
      \put(1078,3418){\makebox(0,0)[r]{\large{ 0.96}}}%
      \put(1078,4253){\makebox(0,0)[r]{\large{ 1}}}%
      \put(1210,484){\makebox(0,0){\large{ 0}}}%
      \put(3074,484){\makebox(0,0){\large{ 500}}}%
      \put(4939,484){\makebox(0,0){\large{ 1000}}}%
      \put(6803,484){\makebox(0,0){\large{ 1500}}}%
      \put(76,2739){\makebox(0,0){\LARGE{$\frac{\sigma_{eff}(r)}{\sigma(0)}$}}}%
      \put(4006,154){\makebox(0,0){\large{$r$ [nm]}}}%
    }%
    \gplgaddtomacro\gplfronttext{%
    }%
    \gplgaddtomacro\gplbacktext{%
      \csname LTb\endcsname%
      \put(2754,2720){\makebox(0,0)[r]{\strut{} 0.9}}%
      \put(2754,3405){\makebox(0,0)[r]{\strut{} 1}}%
      \put(2754,4090){\makebox(0,0)[r]{\strut{} 1.1}}%
      \put(2754,4775){\makebox(0,0)[r]{\strut{} 1.2}}%
      \put(2886,2500){\makebox(0,0){\strut{} 0}}%
      \put(3505,2500){\makebox(0,0){\strut{} 0.02}}%
      \put(4125,2500){\makebox(0,0){\strut{} 0.04}}%
      \put(4744,2500){\makebox(0,0){\strut{} 0.06}}%
      \put(5363,2500){\makebox(0,0){\strut{} 0.08}}%
      \put(2200,3747){\rotatebox{-270}{\makebox(0,0){\large{$\psi^{2}(k;0)$}}}}%
      \put(4124,2170){\makebox(0,0){\large{$ k $ [nm$^{-1}$]}}}%
    }%
    \gplgaddtomacro\gplfronttext{%
    }%
    \gplbacktext
    \put(0,0){\includegraphics{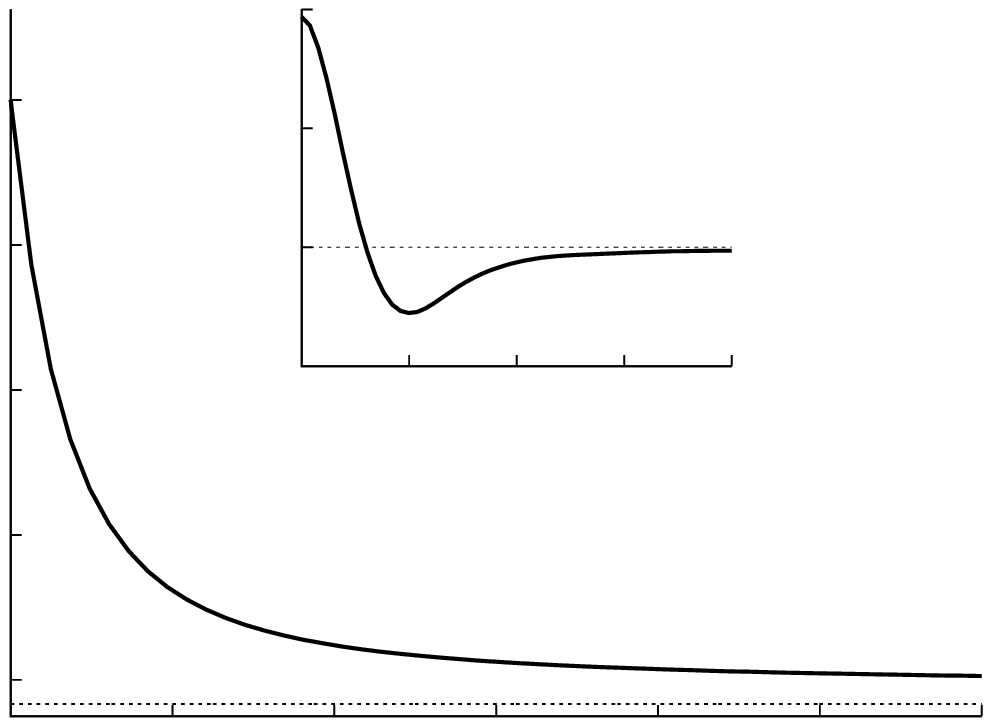}}%
    \gplfronttext
  \end{picture}%
\endgroup
}
\caption{ Numerically obtained  $\sigma_{eff}(r)$ for the conductivity $\sigma(z)$ described in  Fig.~\ref{pot1}. The dotted horizontal line  corresponds to the  asymptotic value  $\sigma(z)$ with $z\to\infty$.  
In the inset $\psi^{2}(k;0)$   is shown.  }
\end{figure}
\begin{figure}
\resizebox{!}{0.30\textwidth}{
\begingroup
  \gdef\gplbacktext{}%
  \gdef\gplfronttext{}%
 \ifGPblacktext
    \def\colorrgb#1{}%
    \def\colorgray#1{}%
  \else
    \ifGPcolor
      \def\colorrgb#1{\color[rgb]{#1}}%
      \def\colorgray#1{\color[gray]{#1}}%
      \expandafter\def\csname LTw\endcsname{\color{white}}%
      \expandafter\def\csname LTb\endcsname{\color{black}}%
      \expandafter\def\csname LTa\endcsname{\color{black}}%
      \expandafter\def\csname LT0\endcsname{\color[rgb]{1,0,0}}%
      \expandafter\def\csname LT1\endcsname{\color[rgb]{0,1,0}}%
      \expandafter\def\csname LT2\endcsname{\color[rgb]{0,0,1}}%
      \expandafter\def\csname LT3\endcsname{\color[rgb]{1,0,1}}%
      \expandafter\def\csname LT4\endcsname{\color[rgb]{0,1,1}}%
      \expandafter\def\csname LT5\endcsname{\color[rgb]{1,1,0}}%
      \expandafter\def\csname LT6\endcsname{\color[rgb]{0,0,0}}%
      \expandafter\def\csname LT7\endcsname{\color[rgb]{1,0.3,0}}%
      \expandafter\def\csname LT8\endcsname{\color[rgb]{0.5,0.5,0.5}}%
    \else
      \def\colorrgb#1{\color{black}}%
      \def\colorgray#1{\color[gray]{#1}}%
      \expandafter\def\csname LTw\endcsname{\color{white}}%
      \expandafter\def\csname LTb\endcsname{\color{black}}%
      \expandafter\def\csname LTa\endcsname{\color{black}}%
      \expandafter\def\csname LT0\endcsname{\color{black}}%
      \expandafter\def\csname LT1\endcsname{\color{black}}%
      \expandafter\def\csname LT2\endcsname{\color{black}}%
      \expandafter\def\csname LT3\endcsname{\color{black}}%
      \expandafter\def\csname LT4\endcsname{\color{black}}%
      \expandafter\def\csname LT5\endcsname{\color{black}}%
      \expandafter\def\csname LT6\endcsname{\color{black}}%
      \expandafter\def\csname LT7\endcsname{\color{black}}%
      \expandafter\def\csname LT8\endcsname{\color{black}}%
    \fi
  \fi
  \setlength{\unitlength}{0.0500bp}%
  \begin{picture}(7200.00,5040.00)%
    \gplgaddtomacro\gplbacktext{%
      \csname LTb\endcsname%
      \put(1210,1286){\makebox(0,0)[r]{\large{ 1}}}%
      \put(1210,2740){\makebox(0,0)[r]{\large{ 1.125}}}%
      \put(1210,4193){\makebox(0,0)[r]{\large{ 1.25}}}%
      \put(1342,484){\makebox(0,0){\large{ 0}}}%
      \put(2434,484){\makebox(0,0){\large{ 200}}}%
      \put(3526,484){\makebox(0,0){\large{ 400}}}%
      \put(4619,484){\makebox(0,0){\large{ 600}}}%
      \put(5711,484){\makebox(0,0){\large{ 800}}}%
      \put(6803,484){\makebox(0,0){\large{ 1000}}}%
      \put(76,2739){\makebox(0,0){\LARGE{$\frac{\sigma_{eff}(r)}{\sigma(0)}$}}}%
      \put(4072,154){\makebox(0,0){\large{$r$ [nm]}}}%
    }%
    \gplgaddtomacro\gplfronttext{%
    }%
    \gplgaddtomacro\gplbacktext{%
      \csname LTb\endcsname%
      \put(4194,1551){\makebox(0,0)[r]{\strut{} 0.8}}%
      \put(4194,2007){\makebox(0,0)[r]{\strut{} 0.9}}%
      \put(4194,2464){\makebox(0,0)[r]{\strut{} 1}}%
      \put(4194,2920){\makebox(0,0)[r]{\strut{} 1.1}}%
      \put(4326,1240){\makebox(0,0){\strut{} 0}}%
      \put(5317,1240){\makebox(0,0){\strut{} 0.02}}%
      \put(6308,1240){\makebox(0,0){\strut{} 0.04}}%
      \put(3700,2235){\rotatebox{-270}{\makebox(0,0){\large{$\psi^{2}(k;0)$}}}}%
      \put(5964,1020){\makebox(0,0){\large{$ k $ [nm$^{-1}$]}}}%
    }%
    \gplgaddtomacro\gplfronttext{%
    }%
    \gplbacktext
    \put(0,0){\includegraphics{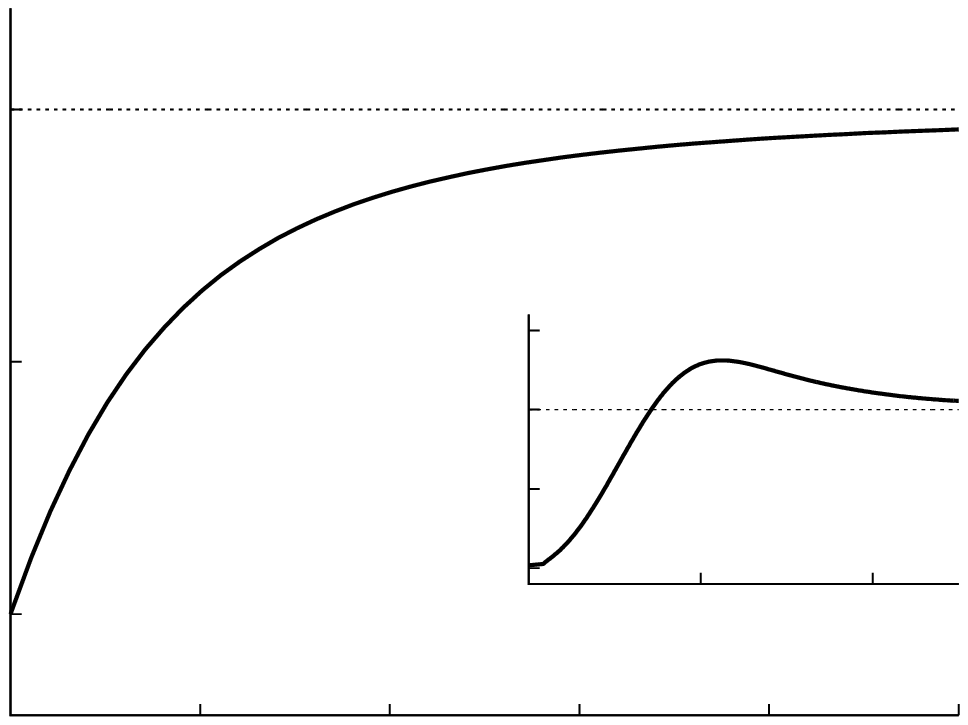}}%
    \gplfronttext
  \end{picture}%
\endgroup
}
\caption{ Numerically obtained  $\sigma_{eff}(r)$ for  the conductivity $\sigma(z)$ described in  Fig.~\ref{pot2}. The dotted horizontal line  corresponds to the asymptotic value of the bulk conductivity.  
In the inset $\psi^{2}(k;0)$  is shown. }
\end{figure}

\section{Beyond the point-source approximation}\label{finitesize}
A more realistic model of the contact supplying the current to the sample involves an area on the surface, where both the function $\Phi$ and the current $\sigma(0)\partial_{z}\Phi$ are given.  This corresponds to the physical situation in which the potential and current at the contact are determined  by the supplying  electrodes. A detailed analysis of the problem goes beyond the scope of the present article.  While the experimental data were acquired with different tips, yielding different contact areas and shapes, the results seem not to depend on them. This is why we will be satisfied with only a few general remarks.

The requirement that for a given area   $\Phi(\mathbf{r},0)$ be equal to a predefined function, can be satisfied by making use of the obtained Green function. For a qualitative discussion one can formulate an analogue of the multipole expansion  known in electrostatics to arrive at quickly vanishing $r^{-n}$ ($n>1$)  terms.

To  take into account the incoming current profile,  further developments are needed. We observe that the current density $\mathbf{j}$ corresponds, upon transformation~(\ref{transf}), to the quantity
\begin{displaymath}
\mathbf{j}=\sqrt{\sigma}\nabla\xi-\frac{\xi}{2\sqrt{\sigma}}\nabla\sigma.
\end{displaymath}
Assuming  boundary condition~(\ref{bound2}) we can draw an analogy to the quantum mechanics and interpret the current density at the surface as the momentum of the incoming particles. This makes it clear that for different $\mathbf{j}$ we can get different penetration depths, and hence different solutions.  
The Green function is built as a weighted sum of all related eigenfunctions, the larger the eigenvalue (energy) the smaller the weight. Thus, the point-source formula is dominated by low-energy features. 
As we have observed, it is  the long-ranged behavior ($r\to\infty$) that is  governed by small eigenvalues ($k\to 0$). Regions that are distant  enough  should not be  affected either by source finite-size effects or by the value of the current density.

\section{Conclusions}
Our experimental data show a systematic change of the surface conductivity with the distance between the probes. To understand the results we have investigated the classical current flow equation. We have developed a theoretical scheme making it evident that the change of the conductivity beneath the surface  impacts the results obtained on the surface, i.e. the well-known formula for the current point-source $\phi\sim 1/\sigma r$ is to be transformed to the  form $\phi\sim 1/\sigma_{eff}(r)r$. Asymptotic  values of $\sigma_{eff}$ have been found to be the actual $\sigma(0)$ value close to the source and the bulk value far from the source. This scheme eases the numerical effort needed to obtain the surface conductivity profile. Furthermore, it delivers a framework to discuss and classify different effects obtained in numerical studies such as that in Ref.~\onlinecite{Hoffmann}.  We have confined our considerations to small deviations of the conductance parameter. 

There are several interesting open questions mentioned in the text, however three  issues are of practical interest. First, the inverse problem, i.e. the existence  of a scheme of surface measurements allowing for reconstruction of the conductivity profile. Second,  capturing the mechanism behind the dimensional reduction~\cite{Hoffmann,APL} seems both interesting and feasible. Third, apparently the condition $\tfrac{d}{dz}\sigma(0)=0$ corresponds to physical observations. It is at odds with the way (envelope) wavefunctions are modeled near the surface~\cite{King}. This contradiction needs to be clarified.

\section{Acknowledgements}
Founding for this research has been provided by EC under the Large-scale Integrating Project in FET Proactive of the 7th FP entitled ``Planar Atomic and Molecular Scale devices" (PAMS). The research was carried out with  equipment purchased thanks to the financial support of the European Regional Development Fund in the framework of the Polish Innovation Economy Operational Program (contract no. POIG.02.01.00-12-023/08).
\appendix
\section{Calculation Details }\label{AppA}
Here we describe in more detail the calculations behind the results outlined in  sect.~\ref{res}. To begin with, we write explicitly  eq.~(\ref{defL}) which is to be solved. 
\begin{equation}\label{app1}
\left(\partial_{x}^{2}+\partial_{y}^{2}+\partial_{z}^{2}+ V(z)\right)\xi=\delta(\mathbf{r}-\mathbf{r}'),
\end{equation}
where the potential is calculated from the conductivity function
\begin{displaymath}
V(z)=\sigma^{-1/2}(z)\frac{d^{2}}{dz^{2}}\sqrt{\sigma(z)}.
\end{displaymath}
The solution of  eq.~(\ref{app1}) defines the Green function for the differential operator. There are many methods of finding Green functions, from which   spectral decomposition~\cite{Green1, Green2} is most appropriate to us. We observe that the related eigenproblem is solved via
\begin{equation}
\hat{L}e^{i \left(p x+q y\right)}\psi(k;z) = \left(p^{2}+q^{2}+k^{2}\right) e^{i \left(p x+q y\right)}\psi(k;z),
\end{equation} 
where we follow notation introduced in sect.~\ref{res}. Both $p$ and $q$ are any real numbers, while $k$ may be considered positive.\\
The Green function may  be expressed as
\begin{equation}
\begin{split}
G(x,y,z;x',y',z')= \\
\int_{0}^{\infty}dk \int_{-\infty}^{\infty} dp\int_{-\infty}^{\infty} dq \frac{e^{i p(x-x')+i q(y-y')}\psi(k;z)\bar{\psi}(k;z')}{p^{2}+q^{2}+k^{2}},
\end{split}
\end{equation}
we consider real $\psi$, so $\psi=\bar{\psi}$. The eigenmodes have to obey a uniform  normalization condition, e.g.
\begin{displaymath}
\int_{0}^{\infty} dz \ \psi(k;z) \psi(k';z)=\delta(k-k'). 
\end{displaymath}

Using  relation A5 from Ref.~\onlinecite{k0}
\begin{displaymath}
\frac{1}{2\pi}\int_{-\infty}^{\infty}\int_{-\infty}^{\infty} \frac{e^{i\mathbf{\xi}\cdot \mathbf{x}}}{|\mathbf{\xi}|^{2}+|\beta|^{2}}d\xi_{1}d\xi_{2}=K_{0}(|\beta||\mathbf{x}|)
\end{displaymath}
we can integrate out $p$ and $q$ to arrive at 
the most important formula in this paper
\begin{equation}\label{green1}
G(x,y,z;x',y',z')=2 \pi \int_{0}^{\infty}d k K_{0} \left(k r \right) \psi(k;z)\psi(k,z').
\end{equation}
where $r=\sqrt{(x-x')^{2}+(y-y')^{2}}$ and $K_{0}$ stands for the modified Bessel function of the second kind. To proceed, the functions $\psi$ and related eigenvaues are needed. This can be done for the potential-free case as shown below (see Appendix~\ref{AppC}).

With no harm to generality we set $x'=0$ and $y'=0$. As the point source is to be located at the surface, we  also set $z'=0$. We look for the surface profile of the Green function, hence we can also  put $z=0$. The function $G(x,y,0;0,0,0)$ corresponds to the sought-after surface profile $\phi(r)$. Now we can argue for the asymptotic formulas  given in sect.~\ref{res}. 
The integral to solve is
\begin{equation}
C= \int_{0}^{\infty}d k \ K_{0} \left(k r \right) \psi^{2}(k;0).
\end{equation}
Substituting $u=kr$, the integral becomes
\begin{equation}
C= \frac{1}{r}\int_{0}^{\infty}d u \ K_{0} \left(u \right) \psi^{2}\left(\frac{u}{r};0\right).
\end{equation}
We note that the Bessel function $K_{0}$ has a logarithmic divergence at the origin and exponentially decays for large arguments. Integrating $K_{0}$ gives
\begin{displaymath}
\int_{0}^{\infty}d u \ K_{0} \left(u \right) = \frac{\pi}{2},
\end{displaymath}
see 11.4.23 in Ref.~\onlinecite{abra}.   As the function $\psi(k;0)$ is bounded, for any desired accuracy $\epsilon>0$ one can find a value $u_{0}$, for which
\begin{equation}\label{ineq}
\left |\left( \int_{0}^{\infty} - \int_{0}^{u_{0}} \right) d u \ K_{0} \left(u \right) \psi^{2}\left(\frac{u}{r};0\right)\right|<\epsilon.
\end{equation}
As a consequence, we can switch to integration over a finite interval.

First we deal with the case of small values of $r$. To this end, we observe that for sufficiently large $k$ the potential may be considered a small perturbation of the eigenstate obtained for the potential-free equation. As such, $\psi(k,z)\approx \cos kz$ and $\psi(k,0)\to 1$.
Hence, there is an argument $t_{\infty}$ such that for any $t>t_{\infty}$ it holds that $\psi(t,0)\approx 1$. As such,  we can split integration into two parts,
\begin{equation}
r C= \int_{0}^{t_{\infty}r}d u \ K_{0} \left(u \right)\psi^{2}\left(\frac{u}{r};0\right)+\int_{t_{\infty}r}^{u_{0}}d u \ K_{0} \left(u \right)\cdot 1.
\end{equation}
The first term vanishes due to  shrinking of the integration interval with $r$ and finally we are left with the limiting value
\begin{equation}\label{asymSmallR}
 C=\frac{\pi}{2 r}.
\end{equation}
Now, we address large values of $r$. We assume that the limit 
\begin{displaymath}
\lim_{t\to 0}\psi(t;0)=\psi(0;0)
\end{displaymath} 
is finite and does not vanish. Hence, there is again $t_{0}$ such that for any $t<t_{0}$ the difference  $\psi(t;0)-\psi(0;0)$ can be neglected. Then
\begin{equation}\label{smallt}\begin{split}
 r C= \int_{0}^{t_{0}r}d u \ K_{0} \left(u \right)\psi^{2}\left(0;0\right)+\\
+ \int_{t_{0}r}^{\min\{u_{0},t_{0}r\}}d u \  K_{0} \left(u \right)\psi^{2}\left(\frac{u}{r};0\right).
 \end{split}
\end{equation}
Now it is the second term which vanishes due to reduction of the integration interval. As such we arrive at the limiting value
\begin{equation}
 C= \frac{\pi \psi^{2}\left(0;0\right)}{2 r}.
\end{equation}
It is easy to note, that eq.~(\ref{Schrod1}) has an zero eigenvalue corresponding to the function $\sqrt\sigma$. This function is related to the freedom left by eq.~(\ref{current1}), which admits shifting its solutions by any constant. The second independent solution $\psi_{0}(z)$ assuming $k=0$ reads 
\begin{displaymath}
\psi_{0}(z)=\sqrt{\sigma(z)}\int_{0}^{z}\frac{dy}{\sigma(y)}.
\end{displaymath}
It vanishes for $z=0$ but has a non-vanishing first derivative.
Following the theorems on continuity of solutions of a differential equation with respect to a parameter~\cite{ode} we make use of the solution $\sqrt{\sigma}$ to arrive at 
\begin{equation}
\psi(0;0)=\sqrt{\frac{\sigma(0)}{\sigma(\infty)}},
\end{equation}
where the denominator appears due to the proper normalization.

To complete the discussion, we address the case, where $\psi(t;0)\sim t$ for $t\to 0$ and $m>0$ (see eq.~(\ref{faza})). In such a case we arrive at the following relation
\begin{equation}\begin{split}
r C= \frac{1}{r^{2}}\int_{0}^{\tilde{t}_{0}r}d u \ u^{2} K_{0} \left(u \right) +\\ +\int_{\tilde{t}_{0}r}^{\min\{u_{0},\tilde{t}_{0}r\}}d u \  K_{0} \left(u \right)\psi^{2}\left(\frac{u}{r};0\right),
\end{split}
\end{equation}
where $\tilde{t}_{0}$ stands for the maximal argument, for which the linear approximation may be considered exact. As a consequence, we obtain the relation 
\begin{equation}
 C= \frac{1}{r^{3}}\int_{0}^{\infty}d u \ u^{2} K_{0} \left(u \right) 
\end{equation}
with a different asymptotic current behavior. The numerical value of the integral above yields
\begin{displaymath}
\int_{0}^{\infty}d u \ u^{2} K_{0} \left(u \right) = \frac{\pi}{2},
\end{displaymath}
as can be interfered  from formula~11.4.22 in~Ref.~\onlinecite{abra}. The case with a negative value of $m$ is more complicated. The spectrum  then has an isolated eigenfunction $\sim \exp{m z}$ with a negative eigenvalue~\cite{adjoint} and hence the Green function becomes more complicated than the functions considered here. \\
It is evident from the numerical examples described in sect.~\ref{res}, that $\sigma_{eff}$ changes on a slightly  larger scale than the potential. To   remark on finding  solutions of eq.~(\ref{Schrod1}) for small eigenvalues $k$, we denote the range of the potential $V$ with $d$. It corresponds to the distance from which the potential may be neglected. Then, in the basis of $\{\cos kz\}_{k>0}$ the matrix element $V_{k,p}$ reads
\begin{equation}
V_{k,p}=\frac{1}{2}\int_{0}^{\infty}dz \ V(z)\left[\cos{(k+p)z}+\cos{(k-p)z}\right].
\end{equation}
For small values of $k$ and $p$ the integral tends to the value
\begin{equation}
V_{k,p} \underset{k,p\to 0} {\longrightarrow}\int_{0}^{d}dz \ V(z),
\end{equation}
if in the range $(0,d)$ functions $\cos(\cdot)$  can be considered constant. As such, the vectors $k$ obeying the inequality
\begin{displaymath}
kd\lesssim 0.1
\end{displaymath}
 may be considered small. This hints at the length scale needed to solve the problem  numerically -- to explore the lowest eigenvalue sector  one needs to work on segments $\sim 100 d$ in length. We used the standard Mathematica~10 routines to diagonalize the operator $\hat{L}$ (with higher precision). The segment lengths that we explore are about 20~000 nm and 40~000 nm. The lowest eigenvalues turned numerically unstable. This is why we excluded several lowest eigenvalues. $\psi^{2}(0,0)$ was found by  extrapolation a function  obtained as quadratic interpolation of $\psi^{2}(k,0)$ for several lowest remaining eigenvalues. As shown in Fig.~6 and 7, the results  are close to the exact values $\sigma(0)/\sigma(\infty)$. A more credible and efficient approach to finding $\psi(k;0)$ is to solve  eq.~(\ref{Schrod1}) for different values of $k$ with the boundary condition~(\ref{bound2}) on a segment larger than the range of the potential $V$. The normalization requires then, that there is a given amplitude of oscillations for all solutions $\psi(k,z)$ in the potential-free region. 

\section{Green functions for the potential-free equation}\label{AppC}
For the potential-free case the eigenfunctions for the eigenvalues $k^{2}$ values yield  $\cos (k z)$ if condition~(\ref{bound2}) is to be satisfied and $\cos\left( k z+\varphi(k)\right)$ if condition~(\ref{bound3}) holds. The phase shift $\varphi$ is given by the formula
\begin{equation}\label{faza}
\varphi=-sign(m)\arccos\frac{k}{\sqrt{k^{2}+m^{2}}},
\end{equation}
where $m$ denotes $\partial_{z} \sigma(0)/2\sigma(0)$ and $sign(m)$ is the sign of $m$. 
For eigenfunctions $\cos(k z)$ we can complete the integration in eq.~(\ref{green1}) to arrive at the following formula
\begin{equation}\begin{split}
G(x,y,z;x',y',z')=\\ 2 \pi \int_{0}^{\infty}d k K_{0}\left(k r \right) 
 \frac{1}{2}\left[\cos k\left(z-z'\right)+ \cos k\left(z+z'\right)\right].
 \end{split}
\end{equation}
Using relation 11.4.14 in Ref.~\onlinecite{abra} we obtain
\begin{equation}\begin{split}
G(x,y,z;x',y',z')=\frac{\pi^{2}}{2} \left(\frac{1}{\sqrt{r^{2}+(z-z')^{2}}}+\right. \\ 
\left. +\frac{1}{\sqrt{r^{2}+(z+z')^{2}}}\right),
\end{split}
\end{equation}
which reduces to the familiar formula $r^{-1}$ at the surface. The term with $z+z'$ appears due to the boundary conditions to ensure that no current escapes through the surface. Similar Green functions are known in  electrostatics to give rise to image charges.

We note that eigenfunctions $\cos kz$  converge to a constant in the limit $k\to 0$, which solves  potential-free eq.~(\ref{Schrod1}) with $k=0$.  This is a demonstration  of theorems on the continuity of solutions of a differential equations  with respect to parameters. 

\section{Experimental data revisited}\label{AppB}
The insight gained from the above analysis of the current flow equation prompts a few comments on the experimental data presented in sect.~\ref{Exp}. The bulk value of the conductivity appears different  for the two samples investigated and reads about $1.4-1.5 \ \Omega^{-1}$cm$^{-1}$ and $1.2-1.3 \ \Omega^{-1}$cm$^{-1}$ for sample A and B, respectively. In both cases it is below the nominal value. The experimental conductivity decreases with the distance from the source, see Fig.~1.  The trend is reproduced in Fig.~6, which suggests an enhancement of the conductivity at the surface. For the p-type doped samples the surface Fermi level is located close to the valence band and hence the formation of an accumulation layer  consistently explains the data. \\
In Fig.~6 and~7 we observe, that the first derivative of $\psi^{2}(k;0)$ vanishes for $k\to 0$. As such, the following asymptotic expansion for large $r$ 
\begin{equation}\label{seriesSigma}
\sigma_{eff}(r)=\sigma(\infty)\left(1-\partial_{k}^{2}\psi^{2}(0;0) \frac{\sigma(\infty)}{\sigma(0)}\frac{1}{2 r^{2}}+\ldots\right)
\end{equation}
is natural. We stop at the first correction term and rewrite the above formula using an effective parameter $\sigma_{1}$
\begin{equation}\label{phenomenology}
\sigma_{eff}(r)=\sigma(\infty)+\frac{\sigma_{1}}{r^{2}}.
\end{equation}
One should bear in mind, that the  terms neglected in eq.~(\ref{seriesSigma}) can be important if the term with  $\sigma_{1}$ becomes relevant. Hence, some caution is needed when attributing a physical meaning to $\sigma_{1}$.\\
To apply the expression~(\ref{phenomenology}) to the experimental setup described in Ref.~\onlinecite{APL} we put the point current source at the point $(0,0,0)$ and the drain at $(D,0,0)$ and calculate the voltage drop between probes located at $\mathbf{x}_{1}=(D\tfrac{1-s}{2},0,0)$ and $\mathbf{x}_{2}=(D\tfrac{1+s}{2},0,0)$. Here the parameter  $0<s<1$ corresponds to the parameter $x$ in Ref.~\onlinecite{APL}. The resulting formula for the measured resistance $R(D;s)$, i.e. the voltage drop between $\mathbf{x}_{1}$ and $\mathbf{x}_{2}$ divided by the current supplemented to the sample, is poorly informative. However, it is interesting to expand it in powers of $1/D$
\begin{equation}\label{ResistanceSeries}
R(D;s)=\frac{s}{\pi \sigma(\infty)(1-s^{2})}\frac{1}{D}-\frac{4 \sigma_{1}s (s^{2}+3)}{\pi \sigma^{2}(\infty)(1-s^{2})^{3}}\frac{1}{D^{3}} + \ldots.
\end{equation}
We confine our considerations to the two lowest orders of expansion written above. For $s$ close to zero they are finite, thus the first term (of the order $D^{-1}$)  dominates. At this level,  the scaling behavior characteristic for three dimensions~\cite{APL} may be seen: the quantity  $D\cdot R(D,s)$ does not depend on $D$. In the case of experimental data such a universality is clearly observed for $D$ equal to 8 and 16~$\mu$m, see Fig.~8. For smaller $D$ the second term ($\sim D^{-3}$)  spoils the universality. The reason is, that for  $s$ close to unity, both terms on the right-hand side of eq.~(\ref{ResistanceSeries}) are singular. The second one explodes quicker and dominates for  big enough  $s$, no matter how large $D$ is. As the second term in eq.~(\ref{ResistanceSeries}) is due to the subsurface variation of the conductivity, the deviations from the universal behavior bear information about the subsurface region. 

\begin{figure}[ht]
\resizebox{!}{0.3\textwidth}{
\begingroup
  \gdef\gplbacktext{}%
  \gdef\gplfronttext{}%
  \makeatother
  \ifGPblacktext
    \def\colorrgb#1{}%
    \def\colorgray#1{}%
  \else
    \ifGPcolor
      \def\colorrgb#1{\color[rgb]{#1}}%
      \def\colorgray#1{\color[gray]{#1}}%
      \expandafter\def\csname LTw\endcsname{\color{white}}%
      \expandafter\def\csname LTb\endcsname{\color{black}}%
      \expandafter\def\csname LTa\endcsname{\color{black}}%
      \expandafter\def\csname LT0\endcsname{\color[rgb]{1,0,0}}%
      \expandafter\def\csname LT1\endcsname{\color[rgb]{0,1,0}}%
      \expandafter\def\csname LT2\endcsname{\color[rgb]{0,0,1}}%
      \expandafter\def\csname LT3\endcsname{\color[rgb]{1,0,1}}%
      \expandafter\def\csname LT4\endcsname{\color[rgb]{0,1,1}}%
      \expandafter\def\csname LT5\endcsname{\color[rgb]{1,1,0}}%
      \expandafter\def\csname LT6\endcsname{\color[rgb]{0,0,0}}%
      \expandafter\def\csname LT7\endcsname{\color[rgb]{1,0.3,0}}%
      \expandafter\def\csname LT8\endcsname{\color[rgb]{0.5,0.5,0.5}}%
    \else
      \def\colorrgb#1{\color{black}}%
      \def\colorgray#1{\color[gray]{#1}}%
      \expandafter\def\csname LTw\endcsname{\color{white}}%
      \expandafter\def\csname LTb\endcsname{\color{black}}%
      \expandafter\def\csname LTa\endcsname{\color{black}}%
      \expandafter\def\csname LT0\endcsname{\color{black}}%
      \expandafter\def\csname LT1\endcsname{\color{black}}%
      \expandafter\def\csname LT2\endcsname{\color{black}}%
      \expandafter\def\csname LT3\endcsname{\color{black}}%
      \expandafter\def\csname LT4\endcsname{\color{black}}%
      \expandafter\def\csname LT5\endcsname{\color{black}}%
      \expandafter\def\csname LT6\endcsname{\color{black}}%
      \expandafter\def\csname LT7\endcsname{\color{black}}%
      \expandafter\def\csname LT8\endcsname{\color{black}}%
    \fi
  \fi
  \setlength{\unitlength}{0.0500bp}%
  \begin{picture}(7200.00,5040.00)%
    \gplgaddtomacro\gplbacktext{%
      \csname LTb\endcsname%
      \put(1078,704){\makebox(0,0)[r]{\strut{} 0}}%
      \put(1078,2017){\makebox(0,0)[r]{\strut{} 2000}}%
      \put(1078,3330){\makebox(0,0)[r]{\strut{} 4000}}%
      \put(1078,4644){\makebox(0,0)[r]{\strut{} 6000}}%
      \put(1210,484){\makebox(0,0){\strut{} 0}}%
      \put(2275,484){\makebox(0,0){\strut{} 0.2}}%
      \put(3341,484){\makebox(0,0){\strut{} 0.4}}%
      \put(4406,484){\makebox(0,0){\strut{} 0.6}}%
      \put(5471,484){\makebox(0,0){\strut{} 0.8}}%
      \put(6537,484){\makebox(0,0){\strut{} 1}}%
      \put(176,2739){\rotatebox{-270}{\makebox(0,0){DR(D;s) [$\Omega  \mu$m]}}}%
      \put(4006,154){\makebox(0,0){s}}%
    }%
    \gplgaddtomacro\gplfronttext{%
      \csname LTb\endcsname%
      \put(5816,1537){\makebox(0,0)[r]{D=2 $\mu$m}}%
      \csname LTb\endcsname%
      \put(5816,1317){\makebox(0,0)[r]{D=4 $\mu$m}}%
      \csname LTb\endcsname%
      \put(5816,1097){\makebox(0,0)[r]{D=8 $\mu$m}}%
      \csname LTb\endcsname%
      \put(5816,877){\makebox(0,0)[r]{D=16 $\mu$m}}%
    }%
    \gplbacktext
    \put(0,0){\includegraphics{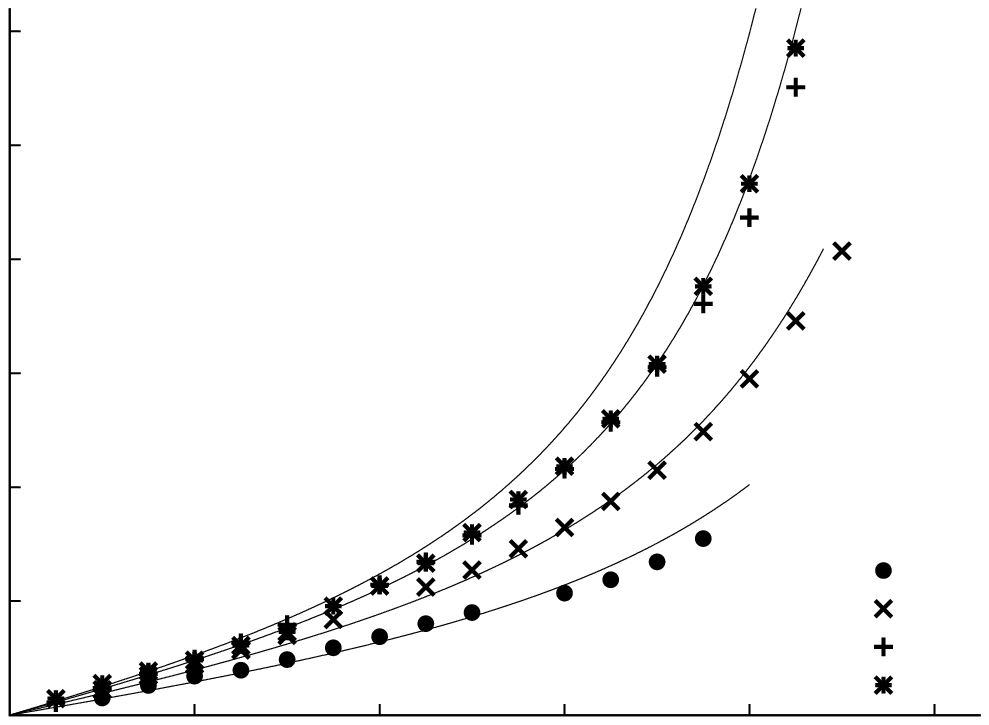}}%
    \gplfronttext
  \end{picture}%
\endgroup
}
\caption{Experimental resistance multiplied by $D$ measured for various $s$ for  sample A  with passivated surface. For $s<0.2$ all the data coincide. The data for $D=8\ \mu$m and $D=16 \mu$m coincide for nearly all values of $s$. Solid lines show numerically obtained data  for the profile $\sigma(z)=\left( 4.5\cosh^{-1}{\tfrac{z}{120}} + 1 \right)\sigma(\infty)$. \label{AGeH} }
\end{figure}
\begingroup
\squeezetable
 \begin{table}[b]
 \begin{tabular}{c||c|c|c||c|c}
 D&\multicolumn{3}{c||}{Sample A}& \multicolumn{2}{c}{Sample B}\\
 \cline{2-6}
 [$\mu m$] & Ge(001) & Ge(001):H & SEM irr. & Ge(001) & Ge(001):H\\
\hline \hline
2  & 1.87 / 8.64 & 1.87 / 8.64 & 2.30 / 3.97 & 1.50 / 1.03 & 1.93 / 2.65 \\
 \hline
4 & 1.73 / 7.53 & 1.73 / 7.53 & 2.05 / 4.97 &  1.66 / 2.90 & 1.55 / 4.34 \\
 \hline
8  & 1.55 / 5.84 & 1.55 / 5.84 & 1.75 / 6.59 & 1.75 / 3.83 & 1.42 / 7.27 \\
 \hline
16 & 1.47 / 17.1 & 1.47 / 17.1  & 1.47 / 12.4 & 1.19 / 40.8 & 1.20 / 22.4 \\
 \hline
\end{tabular}
\caption{$\sigma(\infty)$ / $\sigma_{1}$, i.e. bulk conductivity $\sigma(\infty)$  (in ($\Omega$cm)$^{-1}$)  and $\sigma_{1}$ (in $10^{-14}\Omega^{-1}$cm) from fitting for various surfaces.   The uncertenity for $\sigma(\infty)$ is not larger than 5\% while for $\sigma_{1}$ it varies from several percent for larger $D$ up to 20\% for $D=2\ \mu$m. SEM irr. denotes the surface irradiated by the electron beam from SEM. \label{tab2}}
\end{table} 
\endgroup

The formula assuming~(\ref{phenomenology}) for $R(D;s)$ was used to fit the experimental data for  both samples, see Tab.~\ref{tab2}. As a result, estimated values of $\sigma(\infty)$ vary much less than for a naive fit behind numbers presented in Tab.~\ref{tab}. The values of $\sigma_{1}$ are usually  of the same order, and no clear trend is seen, which is  probably due to the higher order terms. \\
Dimensional analysis  allows to identify a length parameter
\begin{displaymath}
R_{s}=\sqrt{\frac{\sigma_{1}}{\sigma(\infty)}},
\end{displaymath}
a characteristic distance where surface corrections are important.  In our case, it appears about 150-500~nm. This is consistent with the   profile $\sigma(z)$ we apply to reproduce our experimental data, see Fig.~\ref{AGeH}. We use the following function to model conductivity changes 
\begin{equation}
\frac{\sigma(z)}{\sigma(\infty)}=\frac{\tilde{\sigma}}{\cosh{\frac{z}{d}}}+1,
\end{equation}
where $d$, $\tilde{\sigma}$ and $\sigma(\infty)$ are parameters adjusted to the experimental data. This function  satisfies the boundary condition~(\ref{bound2}) and it decays exponentially  except for a region close to the surface. As such, it a reasonable  model describing screening of the surface charge~\cite{Monch}. The parameters $d$ and $\tilde{\sigma}$ have physical meaning while $\sigma(\infty)$ depends on the way one solves the problem (applied normalization condition). As shown in Fig.~\ref{AGeH} simulated data for $d=120$~nm and $\tilde{\sigma}=4.5$ satisfactorily correlate  with the experimental points. Similar results can be obtained for $d$ ranging between 80 and 200~nm  and for $\tilde{\sigma}$ between 3 and 7. Interpretation of these parameters  within the band bending theory, e.g. the Schottky approximation, seems to be misleading as it overestimates the  doping and underestimates the band-bending.

\end{document}